\documentclass{elsart}

\usepackage{epsfig}
\usepackage{amsmath,amssymb,graphicx,caption2,latexsym}
\usepackage[T1]{fontenc}
\usepackage[dvips]{color}

\journal{Physica A}

\newcommand{\bq}{\begin{equation}} 
\newcommand{\eq}{\end{equation}}   
\newcommand{\refm}[1]{(\ref{#1})}  
\newcommand{\difft}[1]{ \frac{d #1}{dt} }

\newcommand{\ve}{{\mathbf e}}

\newcommand{\pa}{\partial}
\newcommand{\N}{{\mathbf N}}
\newcommand{\Nav}{\overline{\N}}

\newcommand{\Om}{\mathbf{\Omega}}
\newcommand{\ksi}{\boldsymbol{\xi}}

\begin{document}

\begin{frontmatter}

\title{Noiseless limit of a ferrofluid ratchet}

\author[ad1]{Volker Becker}, 
\ead{becker@theorie.physik.uni-oldenburg.de} 
\author[ad1]{Andreas Engel} 
\ead{engel@theorie.physik.uni-oldenburg.de} 
\address[ad1]{Institut f\"ur Physik, Carl-von-Ossietzky-Universtit\"at,
     26111 Oldenburg, Germany} 
 
\date{\today} 
 
\begin{abstract} 
The noiseless limit of a thermal ratchet device using ferrofluids is
studied in detail. Contrary to previous claims it is proved that no
directed transport can occur in this model in the absence of
fluctuations. 
\end{abstract} 

\begin{keyword}
thermal ratchet \sep ferrofluids \sep  noiseless limit 
\PACS{05.40.-a, 82.70.-y, 75.50.Mm}
\end{keyword}

\end{frontmatter} 

\section{Introduction}

Rectification of non-equilibrium fluctuations can be accomplished with
the help of so-called {\em ratchets} \cite{AsHa}. In these devices a
periodic potential, i.e. force field with zero spatial average,
and undirected random noise conspire to produce directed transport. Besides
their fundamental importance for statistical mechanics \cite{HaRe}
ratchets have gained renewed interest under the name of ``Brownian
motors'' due to their possible relevance for transport in biological
cells and potential applications in the field of nano-technology
\cite{Linke}. For a comprehensive review of the field see \cite{Reirev}.

Recently a thermal ratchet system using ferrofluids was
introduced \cite{bib:engel1,bib:engel2}. Ferrofluids are colloidal
suspensions of ferromagnetic grains in a suitable carrier liquid
\cite{Ros}. The spatial orientation of the ferromagnetic particles is
influenced by the local vorticity of the flow field of the carrier
liquid as well as by thermal fluctuations due to random collisions
with the molecules of the liquid \cite{Einstein}. Moreover this
orientation can be coupled via the magnetic moment of the grains to
external magnetic 
fields. Choosing a suitable time dependence of this field to drive the
system away from equilibrium it is possible to rectify the
orientational fluctuations of the ferromagnetic particles. More
precisely an external magnetic field {\em without} net rotating
component can be used to set up {\em noise-induced} rotations of the
ferromagnetic grains. Besides other advantages this system has the
attractive feature that by the viscous coupling of the particles to
the surrounding liquid the angular momentum of the many nanoscopic
motors is transferred to the carrier liquid and shows up as a 
{\em macroscopic} torque density which can be easily detected in 
experiments \cite{bib:engel1}.

The necessity of thermal fluctuations for the operation of the ratchet
of the described type in ferrofluids was disputed in \cite{Shcom}. In
fact there are several examples, e.g. so-called {\em rocking} ratchets
(see \cite{Reirev}), in which for appropriate choices of
parameters directed transport may even occur without
fluctuations. However, in order to spin up ferrofluid particles by 
a magnetic field without net rotating component as introduced in
\cite{bib:engel1} the presence of thermal fluctuations is indeed 
{\em indispensable}. This is shown in the present paper where we prove
that in the deterministic dynamics no full rotations of the particles
may occur and that no torque may be transferred from the magnetic
field to the particles. 

To this end we first recall in section II the basic equations for a
single ferromagnetic particle in an oscillating external field as
derived in \cite{bib:engel1}. The ratchet effect was shown to occur in
strongly diluted ferrofluids also such that dipole-dipole interactions
between the ferromagnetic grains may safely be neglected and a
single-particle picture is appropriate for the theoretical
analysis. Setting the noise intensity equal to zero we investigate in
sections III and IV the details of the deterministic  dynamics of the
particle and show that only solutions without full rotation of the
particle are possible. Finally, section V contains the main conclusion.

\section{Basic equations}
We consider a spherical particle of volume $V$ and magnetic moment
$\mathbf{m}$, subject to a time dependent magnetic field of the form 
\bq
\mathbf{H}=\left( H_x,H_y(t),0) \right)
\eq
where $ H_y(t) $ is a periodic function with period $ 2 \pi /
\omega  $. As instructive example \cite{bib:engel1} we will consider
the special case  
\bq
        H_y(t)=\alpha \cos( \omega t) + \beta \sin( 2 \omega t +
        \delta) \label{equ:timedependH} 
\eq 
where $\alpha$, $\beta$ and the phase shift $\delta$ are control
parameters. The particle is immersed in a fluid of viscosity
$\eta$. 

To describe the orientation of the particle we use the
unit vector $ \mathbf{e} = \mathbf{m}/m $ where $\mathbf{m}$ denotes the
the magnetic moment of the particle and $m$ its modulus. Changes of
$\mathbf{e}$ are described by the equation 
\bq
        \difft{ \mathbf{e} } = \Omega \times \mathbf{e}
        \label{equ:changeofe}, 
\eq
where $ \Omega $ is the angular velocity of the particle.

 Furthermore we consider an overdamped stochastic dynamics in which the
 magnetic torque  
\bq
        \mathbf{N}_{mag}=m \mathbf{e} \times \mathbf{H}
        \label{equ:magnettourqe}  
\eq
and the stochastic torque \cite{bib:Coffey}, which results from the
interaction between the particle and the surrounding liquid,  
\bq
        \mathbf{N}_{stoch}=\sqrt{2D}\, \ksi (t)
        \label{equ:stochastictourque} 
\eq     
is counterbalanced by the viscous torque \cite{bib:landauhydro}:  
\bq
     \mathbf{N}_{vis}=-6 \eta V \mathbf{\Omega} \label{equ:viscoustourqe}
\eq
In equation \refm{equ:stochastictourque}, $\ksi(t)$ is a
vector of Gaussian white noise with zero mean and unit variance. The
noise intensity $D$ is related to the temperature $T$ of the liquid by
the Einstein relation: $D=6 \eta V k_B T $, where $k_B$ 
stands for the Boltzmann constant. 
From \refm{equ:magnettourqe}, \refm{equ:stochastictourque} and
\refm{equ:viscoustourqe} we find: 
\bq
        6 \eta V \mathbf{\Omega} =m \mathbf{e} \times \mathbf{H} +
        \sqrt{2D}\, \ksi(t). 
\eq
This relation together with equation \refm{equ:changeofe}  yields a
closed equation for the time evolution of $ \mathbf{e} $ of the form  
\bq
        \difft{\mathbf{e}}=\frac{m}{6 \eta V}\left( \mathbf{e} \times
          \mathbf{H} \right) \times \mathbf{e} + \frac{\sqrt{2D}}{6
          \eta V}\, \ksi \times \mathbf{e} \label{equ:ev1}.  
\eq
Introducing dimensionless units we measure time in units of
the inverse driving frequency, $ t \rightarrow t/ \omega $, and use
$ 6 \eta V \omega / m$ as unit for the magnetic field strength 
$\mathbf{H} \rightarrow (6 \eta V \omega/m)\, \mathbf{H} $. The noise
intensity $D$ is scaled according to $ D \rightarrow \left( 6 \eta V
\right)^2 D $. Eq. \refm{equ:ev1} then reduces to 
\bq
        \difft{\mathbf{e}}=\left( \mathbf{e} \times \mathbf{H} \right)
        \times \mathbf{e} + \sqrt{2D}\,\ksi \times
        \mathbf{e}. \label{equ:evolution}  
\eq

It is convenient to parametrize the orientation of the particle by the
two angles $ \theta $ and $\phi$ according to  
\bq
        \mathbf{e}=\left( \sin \theta \cos \phi, \sin \theta \sin \phi
          , \cos \theta \right) 
\eq
The Langevin equations for the time evolution of these angles are then
of the form \cite{bib:Coffey,bib:raible} 
\begin{align}
        \difft{ \theta}&= -\frac{\partial}{\partial \theta} U + D \cot
        \theta + \sqrt{2D} \,\xi_{\theta}  \label{equ:langevin1} \\ 
        \difft{ \phi}&=-\frac{1}{\sin ^2 \theta}
        \frac{\partial}{\partial \phi} U + \frac{\sqrt{2D}}{\sin \theta} \,
        \xi_\phi . \label{equ:langevin2} 
\end{align}
where we have introduced the potential, 
\bq
U(\theta,\phi)= -\ve\cdot \mathbf{H} = -\sin \theta \left( H_x
  \cos \phi + H_y(t) \sin \phi \right) \label{equ:potential}. 
\eq

The observable of principal interest for the thermal ratchet effect in
ferrofluids is the average torque $\Nav$ arising at the magnetic
particle in the long time limit. Here the average is over time and
hence includes both the ensemble average over different realizations
of the noise as well as the average over the time dependence of the
external magnetic field.  
The focus of the present paper is on the $T \rightarrow 0 $ limit
implying $D\rightarrow 0$. The other system parameters like the fluid
viscosity are assumed to stay constant. We then find for the
averaged torque from the dimensionless forms of
\refm{equ:magnettourqe} and \refm{equ:viscoustourqe}:   
\bq
        \overline{ \mathbf{N}} = \overline{ \ve\times
          \mathbf{H} }=\overline{-\mathbf{\Omega}}
        \label{equ:atourqe2}, 
\eq
since the stochastic torque is zero in the absence of fluctuations.
 
For later use it is instructive also to study the case in which the
particle orientation is confined to the plane defined by 
$\theta\equiv\pi/2$. In this case we have $\Nav=(0,0,\overline{N}_z)$ and 
$\Om=(0,0,\pa \phi/\pa t)$. Therefore from eq.~\refm{equ:atourqe2} we
find 
\bq
        \overline{N_z} = \lim_{(t_2-t_1) \rightarrow{\infty}}
        \frac{1}{t_2-t_1} \int_{t_1}^{t_2} \!\!dt' \;
                          \frac{\pa\phi}{\pa t'}(t')        
\eq
and hence 
\bq  
  \overline{N_z} = \lim_{(t_2-t_1)\rightarrow{\infty}}
  \frac{\phi(t_2)-\phi(t_1)}{t_2-t_1} \label{equ:norot} 
\eq
In the absence of particle rotation, i.e. if $  0 < \phi < 2\pi $,
therefore no average torque may arise. 

In the following two sections we study the deterministic dynamics
given by (\ref{equ:evolution}) with $D=0$ in detail. We first consider
the case $H_x=0$ and then deal with the more general situation where 
$H_x\neq 0$.  


\section{The case $H_x=0$}
In this case the external field has the form    
\bq
\mathbf{H}= \left( 0,H_y(t),0 \right)
\eq
with a general time-dependent $H_y(t)$. For $D=0$
eq.~\refm{equ:evolution} yields the following 
equations for the components of $\mathbf{e}$:
\begin{align}
\difft{e_x} &= - e_x H_y(t) e_y \label{equ:cartx} \\
\difft{e_y} &= - e_y^2 H_y(t) + H_y(t) \label{equ:carty} \\
\difft{e_z} &= - e_z H_y(t) e_y \, . \label{equ:cartz} \\ 
\end{align}
This system of differential equations is to be completed by
appropriate initial conditions at some initial time $t_0$. 
Without loss of generality we can choose the coordinate system in such
a way that $ e_z(t_0)=0 $, i.e. we take the $x$-$z$-plane as the
plane defined by $\mathbf{e}(t_0)$ and the direction of the 
magnetic field. From equation \refm{equ:cartz} then 
follows that $e_z$ is identically zero, $ e_z(t)\equiv 0 $.  

>From \refm{equ:cartx} and \refm{equ:carty} we find 
\bq
        e_y(t)= \tanh \left( \int_{t_0}^t dt' H_y(t') + \mbox{arctanh}
          \, e_y(t_0) \right) \label{equ:hy1} 
\eq
and 
\bq 
e_x(t) = e_x(t_0) \exp\left( - \int_{t_0}^t dt' H_y(t') e_y(t') \right) \label{equ:hx1}
\eq
The integral $I$ in the exponential function in (\ref{equ:hx1}) can be
determined using \refm{equ:hy1}  
\bq 
        I=\int_{t_0}^t dt' H_y(t') \tanh \left( \int_{t_0}^{t'} dt''
          H_y(t'')+\mbox{arctanh} \, e_y(t_0) \right) \, .
\eq
Substituting \mbox{$ u(t)= \int_{t_0}^t dt'
  H_y(t')+\mbox{arctanh}\, e_y(t_0) $} this gives 
\bq
        I=\int_{u(t_0)}^{u(t)} du \tanh(u) = \ln \left(
          \frac{\cosh\left(u(t)\right)}{\cosh\left( u(t_0) \right)}
        \right) 
\eq
and we finally get the solution
\begin{align} 
        e_x(t)&=e_x(t_0) \frac{ \cosh \big(\mbox{arctanh}e_y(t_0)
          \big)}{\cosh \left( \mbox{arctanh}\, e_y(t_0)  + \int_{t_0}^t
            dt' H_y(t')\right)} \label{equ:evx} \\ 
        e_y(t)&=\tanh \left( \int_{t_0}^t dt' H_y(t') + \mbox{arctanh}\,
          e_y(t_0) \right)\, . \label{equ:evy}  
\end{align}

\begin{figure}
\centering
\scalebox{1}{\includegraphics{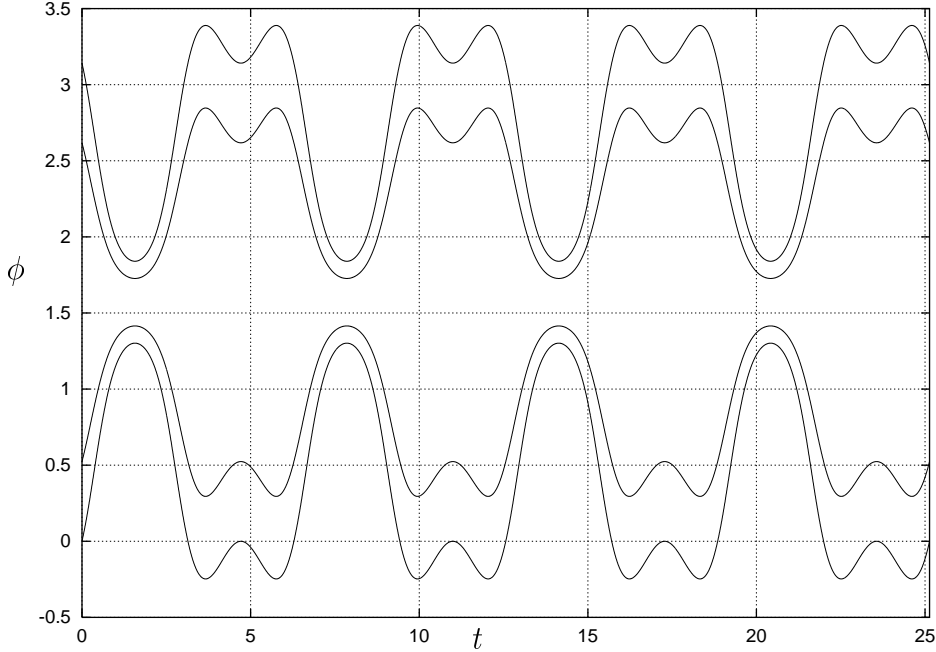}}
\caption{Trajectories $\phi(t)$ for the case $H_x\equiv 0$ and for the
  time dependence of $H_y(t)$ as defined in
  \refm{equ:timedependH} and $t_0=0$. The parameter values are chosen as
  $\alpha=1$, $\beta=1$, and $\delta=0$. The different curves
  correspond to different initial conditions, $\phi_0=\pi\, ,5\pi/6\,
  ,\pi/6$ and $0$ (from top to bottom).   There are no rotary 
  solutions in the absence of fluctuations and hence no average torque
  may arise. Note that the trajectories are different for different 
  initial conditions, there is hence no unique long-time
  behaviour.\label{fig:hx0}} \end{figure} 

>From \refm{equ:evx} it follows that there are no full rotations of the
particle, because the $x$-component of $ \mathbf{e}$ cannot change
sign. Expressed in terms of $ \theta $ and $ \phi $, equations
\refm{equ:evx}, \refm{equ:evy} for this choice of coordinates 
become 
\begin{align}
        \theta (t)& = \frac{\pi}{2} \\
        \phi (t)&= \mbox{arctan} \left( \frac{ \sinh \left( \int_{t_0}^t
              dt' H_y(t') \right) +\cosh \left( \int_{t_0}^t dt' H_y(t')
            \right) \sin \phi_0}{\cos \phi_0} \right),    
\end{align}
where  $ \phi_0=\phi(t_0) $ The domain of the arctan function has 
to be chosen such that $\mbox{arctan}\,(\tan(\phi_0))=\phi_0$. Some
trajectories of $\phi(t)$, for the time dependence $H_y(t)$ as given
in \refm{equ:timedependH} and a certain choice of parameters are shown in
figure \ref{fig:hx0}. Note that there is no unique long time behaviour
for $H_x=0$. The trajectories of $\phi(t)$ are different for different
initial conditions. 

The main conclusion  of this section is the absence of full rotations
of the ferrofluid particle for $H_x=0$. Hence $\phi$ 
is bounded to an interval $ \phi_{min} < \phi < \phi_{max} $. Together
with eq.~\refm{equ:norot} it follows that for $H_x=0$ and any
time dependence $H_y(t)$ there is no average torque.


\section{The case $H_x>0$}
In the previous section we have discussed the behavior of a
ferromagnetic particle in a time-dependent field in $y$-direction and
we have shown that in this case no full rotations of the particle may
occur. Intuitively it is clear that an additional constant field
in $x$-direction should not change this result. Nevertheless the
argument given in the previous section according to which the
$x$-component of the orientation vector $\ve $ cannot 
change sign does no longer hold true. If, e.g., $e_x(t_0)$ is
negative and $H_x$ is positive while $H_y(t)=0$, the
orientation of $\mathbf{e}$ will tend toward the direction of $\mathbf{H}$
and therefore $e_x$ has to change sign. It is therefore necessary to
take a closer look on the possible implications of a constant magnetic
field in $x$-direction for the motion of the magnetic particle. We
will show in this section that even in the presence of such a field
$\theta $ will converge to $\pi/2 $ in the long time limit while
$\phi$ will be confined to an interval  
$ \phi_{min} < \phi < \phi_{max}$.   

We start with eqs. \refm{equ:langevin1} and
\refm{equ:langevin2} in the deterministic limit, $D=0$. Using 
eq.~\refm{equ:potential} they can be written in the form
\begin{align}
  \difft{ \theta}&=\cos \theta \big(H_x \cos\phi + H_y(t) \sin \phi\big)
        \label{equ:dettheta} \\ 
        \difft{ \phi}&=\frac{1}{\sin \theta} H_x
        \big(G(t) \cos \phi - \sin \phi\big) ,  \label{equ:detphi} 
\end{align}
where we have introduced the function $G(t)=H_y(t)/H_x$. 

\begin{figure}
\centering
\scalebox{1}{\includegraphics{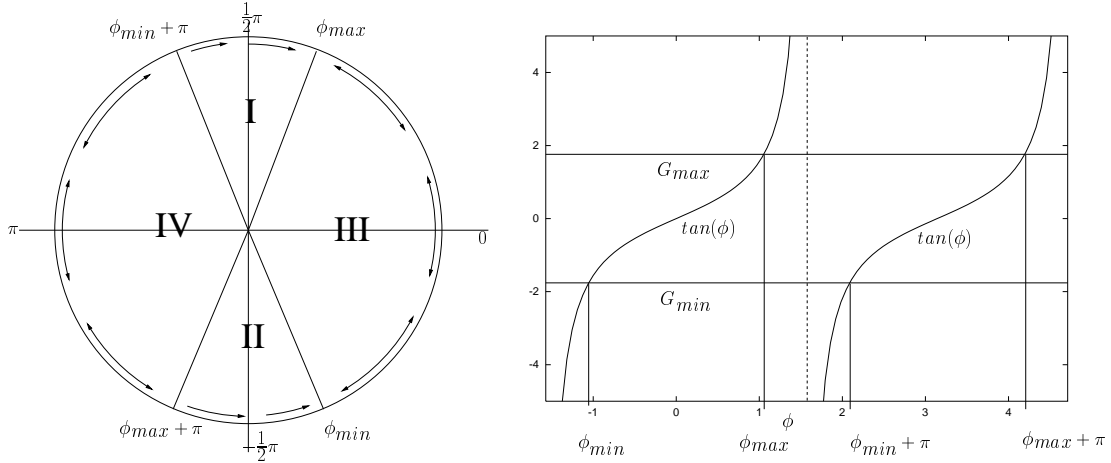}}
\caption{Left: Regions with different values of $d\phi/dt$ as shown by
  the arrows. In region I $d\phi/dt<0$ for all
  $t$, in region II $d\phi/dt>0$ for all $t$. In regions III and IV
  the sign of $d\phi/dt$ depends on the actual value of $G(t)$. 
  Right: Definition of $\phi_{max}$ and $\phi_{min}$ and graphical
  determination of the sign of $d\phi/dt$. \label{fig:phiregion}}
\end{figure}

Let us first look at the sign of $d\phi/dt$ which is given by 
\bq
        \mbox{sgn}\,\left( \difft{\phi} \right) = 
   \mbox{sgn}(\cos\phi) \;\mbox{sgn}(G(t)- \tan\phi). \label{equ:sgn}  
\eq    
Since $G(t)$ is a periodic function it has a maximum $G_{max}$ and a
minimum $G_{min}$. Denote by $\phi_{max}$ and $\phi_{min}$ the
solutions of the equation $\tan\phi=G_{max}$ and $\tan\phi=G_{min}$
respectively in the interval $(-\pi/2,\pi/2)$. Assuming $\phi$ to
belong to the interval $(-\pi/2,3\pi/2)$ it is easy then to show that
$d\phi/dt$ is always (i.e. for all $t$) positive if
$-\pi/2<\phi<\phi_{min}$ or $\phi_{max}+\pi<\phi<3\pi/2$ (region II),
and that it is always negative for $\phi_{max}<\phi<\phi_{min}+\pi$
(region I) (cf. fig.\ref{fig:phiregion}). For the remaining values of
$\phi$ the sign of $d\phi/dt$ depends on the actual value of $G(t)$. 
Note that at $\phi=\pm\pi/2$ both $(G(t)- \tan\phi)$
and $\cos\phi$ change sign such that $d\phi/dt$ does not. Hence these
``critical'' points corresponding to $e_x=0$ belong to regions 
in which the sign of $d\phi/dt$ is independent of time. 

In order to discuss now the time evolution of $\theta$ and $\phi$ we
have to consider different initial conditions for $\phi$. Quite
generally we may assume $H_x>0$ without loss of generality. 
Let us first consider the case $-\pi/2\le\phi_0\le\pi/2$, i.e. 
$\phi$ starts in regions I, II, or III. Then $\phi(t)$ has to reach
region III sooner or later and will not be able to leave it again
(cf. fig.\ref{fig:phiregion}). Strictly speaking this is correct only 
if $\phi(t)$ evolves continuously. But we have to take into account
also that there is the possibility of {\em discontinuous} changes of
$\phi$ when $\theta$ reaches the limiting values $\theta=0$ or
$\theta=\pi$. However, this cannot happen either. To prove this
statement we introduce the quantity  
\bq
        h=\frac{e_z}{e_x} \label{equ:hilf}\, .
\eq
Using (\ref{equ:evolution}) with $D=0$ its time derivative is given by 
\bq
        \difft{h}=-\frac{H_x e_z}{e_x^2}=-\frac{H_x}{e_x}\,h \, ,
\eq
or expressed in terms of $\theta$ and $\phi$
\bq\label{equ:h7}
        \difft{h}=-\frac{H_x}{\cos \phi \sin \theta} \,h \, . 
\eq
Hence 
\bq
        h(t)=h(t_0) \exp \left( -\int_{t_0}^t dt' \frac{H_x}
         {\cos\phi(t') \sin \theta(t')} \right) \label{equ:ehilf} 
\eq
Accordingly,  as long as $\phi$ stays in the interval $(-\pi/2..\pi/2)$
the integral in the exponent of (\ref{equ:ehilf}) grows and
consequently $|h(t)|$ monotonically decreases with time. Therefore 
$\theta$ cannot reach $0$ or $\pi$. 

Hence $\phi(t)$ cannot leave
region III neither by continuous nor by discontinuous changes. On the
one hand this ensures that $\phi_{min}\le\phi(t)\le\phi_{max}$ for all
$t$ on the other hand it implies via (\ref{equ:ehilf}) that $h(t)\to 0$ for
large $t$ and hence that $\theta$ converges asymptotically to $\pi/2$.

If $\pi/2<\phi_0<3\pi/2$ the evolution of $\phi$ starts in regions I,
II, or IV. If at some later time $\phi(t)$ is found in the interval
$(-\pi/2, \pi/2)$ we are back to the previous case. If not and
$h(t_0)\neq 0$ eq.~(\ref{equ:ehilf}) implies that $|h(t)|$ increases
monotonically with time and accordingly $\theta$ tends to either $0$
or $\pi$. By symmetry both cases are equivalent so each other so let us focus
on $\theta\to 0$. Then $h\sim 1/\theta$ and eq.~(\ref{equ:h7}) acquires
the asymptotic form $dh/dt=C\,h^2$ with some positive constant
$C$. Therefore there will be a {\em finite time} singularity in the
solution $h(t)$ and we get $\theta(t_1)=0$ for some finite
$t_1$. Since the magnetic field has a positive $x$-component it is
clear that for $t>t_1$ we will have $-\pi/2\le\phi(t)\le\pi/2$ and
hence we are again back to the first case. 

Summing up, except for a set of measure zero, namely 
$\phi_{min}+\pi<\phi_0<\phi_{max}+\pi$ and $\theta=\pi/2$, all initial
conditions give rise to a long time dynamics with values of $\phi$ between
$\phi_{min}$ and $\phi_{max}$. In any case also for $H_x\neq 0$  
no full rotations of the particle are possible since these would imply
that $\phi(t)$ lies for some $t$ in region III which it were unable to
leave again. As in the case $H_x=0$ we then find from \refm{equ:norot}
that no average torque is transferred from the magnetic field to the
particle or its surrounding liquid.

\section{Conclusion}
By a detailed analysis of the deterministic dynamics of a magnetic
dipole in an external magnetic field with constant $x$-component and
time periodic $y$-component we have proved that 
{\em no full rotations} of the particle may occur. This shows that for
the thermal ratchet effect in ferrofluids reported in
\cite{bib:engel1} thermal fluctuations are {\em indispensable}.


\begin{thebibliography}{99}
\bibitem{AsHa} R. D. Astumian and P. H\"anggi, {\it Physics Today} {\bf
    55}, 33 (2002)
\bibitem{HaRe} S. L. Harvey and A. F. Rex (eds.), 
  {\it Maxwell's Demon: Entropy, Information, Computing} (Adam Hilger,
  Bristol,   1990)
\bibitem{Linke}  H. Linke (ed.) {\it Special issue on ratchets and
    Brownian motors: Basic experiments and applications}, {\it
    Appl. Phys. A} {\bf 75}, 167-352 (2002)
\bibitem{Reirev} P. Reimann, {\it Phys. Rep.} {\bf 361}, 57 (2002) 
\bibitem{bib:engel1} A. Engel, H. W. Müller, R. Reimann, A. Jung, 
  {\it Phys. Rev. Lett.} {\bf 91}, 060602, 2003 
\bibitem{bib:engel2} A. Engel, P. Reimann, {\it Thermal ratchet
    effects in ferrofluids}, to be published in {\it Phys.~Rev.~E.}, 
   {\tt cond-mat/0405393}
\bibitem{Ros} R. E. Rosensweig, {\it Ferrohydrodynamics}, (Cambridge
  University   Press, Cambridge, 1985) 
\bibitem{Einstein} A. Einstein, {\it Ann. d. Physik} {\bf 19}, 371
  (1906)  
\bibitem{Shcom} M. I. Shliomis, {\it Phys. Rev. Lett.} {\bf 92},
  188901 (2004)
\bibitem{bib:landauhydro} L. D. Landau and E. M. Lifshitz, {\it Fluid
  Mechanics}, 2nd ed., \S 20, (Pergamon, New York, 1984)  
\bibitem{bib:Coffey} W. T. Coffey, Y. P. Kalmykov, and J. T. Waldron,
  {\it The Langevin equation with applications in physics, chemistry
    and electrical engineering}, (World Scientific, Singapoure, 1996) 
\bibitem{bib:raible} M. Raible and A. Engel, {\it
    Appl. Organometal. Chem.} {\bf 18}, 536 (2004)   
\end{thebibliography}
\end{document}